\documentstyle[twoside,fleqn,espcrc2]{article}

\def\epsfpreprint{Y}   

\if \epsfpreprint Y \input epsf \fi
   \def\fhi{\vec \phi}
    \def\cm{~~,}

\def\np{Nucl. Phys.~}
\def\pl{Phys. Lett.~}

\newcommand{\AmS}{{\protect\the\textfont2
  A\kern-.1667em\lower.5ex\hbox{M}\kern-.125emS}}

\def\captionA{Leading order cutoff effects in the invariant $\pi-\pi$
scattering amplitude at $90^0$ for the na\"\i ve actions (lines on the
left) and for the actions with a four derivative term turned on to
maximal allowed strength (lines on the right). The dotted line
represents center of mass energies $W=2M_H$, the dashed line $W=3M_H$
and the solid line $W=4M_H$. The PV case here involves a
$(\partial^2)^3$ term.}

\def\captionB{Leading order cutoff effects in the invariant $90^0$
$\pi-\pi$ scattering amplitude vs. $\beta_2$ ($\beta_2$ is
proportional to $b_1$) for three values of $M_H$ on the $F_4$ lattice.
The center of mass energy is set at $W=2M_H$.}

\def\captionC{The regularization independent part of the width vs.
$M_H$. The solid line displays the large $N$ result scaled to $N=4$
and the dotted line shows the leading order term in perturbation
theory.}

\def\captionD{The Higgs mass in units of $f_\pi$ versus the Higgs mass
in lattice units. The solid line is the large $N$ result and the
diamonds the numerical result for the naive action. The dotted line is
the large $N$ result and the squares the numerical result for the
action with the four derivative term turned on to maximum allowed
strength}

\def\captionE{The energy spectrum for various lattice sizes.}

\title{Regularization dependence of the Higgs mass triviality bound.
\thanks{Speakers: P. Vranas on large $N$ analysis and M. Klomfass on
numerical analysis.}}
\author{~Urs M. Heller,\address{SCRI, The Florida State University,
Tallahassee, FL 32306-4052, U.S.A.}
Markus Klomfass,\address{Physics Department,
Columbia University, New York, NY 10027, U.S.A.}
Herbert Neuberger,\address{Department of Physics and Astronomy,
Rutgers University, Piscataway, NJ 08855, U.S.A.}
and
Pavlos M. Vranas $^{\rm a}$ }

\begin{document}

\begin{abstract}
We calculate the triviality bound on the Higgs mass in scalar field
theory models whose global symmetry group $SU(2)_L \times SU(2)_{\rm
custodial} \approx O(4)$ has been replaced by $O(N)$ and $N$ has been
taken to infinity.  Limits on observable cutoff effects at four
percent in several regularized models with tunable couplings in the
bare action yield triviality bounds displaying a large degree of
universality. Extrapolating from $N=\infty$ to $N=4$ we conservatively
estimate that a Higgs particle with mass up to $0.750~TeV$ and width
up to $0.290~TeV$ is realizable without large cutoff effects,
indicating that strong scalar self interactions in the standard model
are not ruled out. We also present preliminary numerical results of
the physical $N=4$ case for the $F_4$ lattice that are in agreement
with the large $N$ expectations.
\end{abstract}

\maketitle

\section{Large $N$ analysis}

We examine the regularization scheme dependence of the triviality
bound on the Higgs mass in a $O(N)$ symmetric scalar field theory to
leading order in $1/N$ \cite{LRGN}.  Our purpose in doing this is
two-fold: we wish to investigate the issue by explicit, analytical
calculations and we need the results both directly and indirectly to
complement numerical work at $N=4$.  The results are needed directly
for estimating cutoff effects on physical observables that are not
accessible by Monte Carlo and indirectly for guiding our search in the
space of lattice actions to the region where heavier Higgs particles
are possible.

A preliminary account of some of our results has been presented in
\cite{{PLBHNV},{LATT91},{ROME}}.  Our general reasons for suspecting
that the present numbers for the bound are too low in the context of
the minimal standard model have been explained before
\cite{{PLBHNV},{TALLA},{CAPRI}} and will not be repeated here.

The basic logic of our approach \cite{{CAPRI},{BBHN1},{PLBF4}} is that
all possible leading cutoff effects can be induced by adding only a
few higher dimensional operators with adjustable coefficients to the
standard $\lambda \Phi^4$ action. We find that the highest Higgs
masses are obtained in the nonlinear limit (infinite $\lambda$
limit).  This leads us to consider nonlinear actions with the leading
cutoff effects induced by four derivative terms with tunable
couplings.

We use one class of continuum regularization schemes, of Pauli Villars
type (PV), and three kinds of lattice regularizations: $F_4$,
Hypercubic (HC), and Symanzik Improved Hypercubic (SI).  For these
regularization schemes the action, when expanded for slowly varying
fields to order momentum to the fourth power, is of the form
\begin{eqnarray}
S_c&=&\int_x ~ \left[ {1\over 2} \vec \phi (-\partial^2
+2 b_0 \partial^4 )\vec \phi
-{b_1 \over {2N}} (\partial_\mu \vec \phi
\cdot \partial_\mu \vec \phi )^2 \right.\nonumber\\
&-&\left. {b_2 \over {2N}} (\partial_\mu \vec
\phi \cdot \partial_\nu \vec \phi - {1\over 4} \delta_{\mu , \nu }
\partial_\sigma \vec \phi \cdot \partial_\sigma \vec \phi )^2 \right],
\label{eq:contaction}
\end{eqnarray}
where $\fhi^2 = N\beta$. There are four control parameters in this
action but one is redundant since to this order it can be absorbed
into the other parameters by a field redefinition. We choose to absorb
the parameter $b_0$ into $b_1$ and $b_2$. However, eliminating the
dependence in $b_0$ to order momentum to the fourth power in the bare
action does not necessarily imply that the dependence in $b_0$ has
been eliminated to leading order in the inverse cutoff in the physical
observables. The vacuum fluctuations are ``aware'' of the full bare
action and will carry that information to the non universal part of
the physical observables. Still, the remaining effect of $b_0$ is probably
small and, as far as the value of the bound is concerned,
one may be able to cover the whole range of leading cutoff effects by
varying $b_1$ and $b_2$ only. As we will see, our results indicate
that this is realized to a good extend, leading to an approximate
universality of the Higgs mass triviality bound.

For PV we simply set $b_0=0$, and for the lattice regularizations
$b_0$ is set to the na\"\i ve value obtained in the expansion of the
lattice kinetic energy term.  We calculate the phase diagram for each
regularization and find a second order line that ends at a tricritical
point where a first order line begins.  We study the physically
interesting region close to the second order line in the broken phase.
The parameter $\beta$ corresponds to the relevant direction and is
traded, as usual, for the pion decay constant $f_\pi$. The parameters
$b_1$ and $b_2$ control the size of the leading order cutoff effects
for a given value of $f_\pi$. Our analysis shows that, to this order,
the quantities we consider do not depend on the parameter $b_2$.
Therefore, a very simple situation emerges with the scale set by
$f_\pi$ and the leading cutoff effects parametrized by only one
parameter $b_1$.  This simplification at infinite $N$ makes it easy to
relate very different regularization schemes, and leads to a
reasonably ``universal'' bound on the renormalized charge $g$ and
therefore on the Higgs mass.

We calculate the leading cutoff effects on two physical quantities,
namely the width of the Higgs particle and the $90$ degrees
$\pi-\pi$ scattering cross section. We find that they are given by
the product of a ``universal'' factor (identical for all
regularizations considered) which only depends on $g$ and the
dimensionless external momenta, and a non--universal factor that
depends on the parameter $b_1$ but not on $g$ or the momenta. The
universal factor associated with the width is different from the one
associated with the cross section, but the non--universal factors are
identical for a fixed regularization scheme. These properties suggest
that, to leading order in $1/N$, all cutoff effects in on--shell
dimensionless physical quantities (viewed as functions of dimensionless
momenta) are given by an effective renormalized action,
\begin{equation}
S_{eff}=S_R
+ c \exp [-{{96\pi^2 }\over g} ] {\cal O} \cm
\label{eq:effact}
\end{equation}
where
${\cal O}$ is a renormalized operator, $c$ is a $g$ independent free
parameter containing all the non--universal information (i.e. a
function of $b_1$ only), and $S_R$ is describing the usual universal
part of physical observables with the unit of energy set by $f_\pi =1$.
In the general RG framework this representation of $S_{eff}$ is not
unreasonable if one accepts that at $N=\infty$ the number of
independent operators that contribute to observables at order
$1/\Lambda^2$ decreases by one relatively to $N<\infty$. Because of the
nonlinear relationship between $c$ and the parameters in the bare
action, the range in which $c$ is allowed to vary depends on the type
of action chosen. Different actions that realize the same $c$ are
indistinguishable to order $1/\Lambda^2$; all actions have large
regions of
overlap but the triviality bounds are obtained from
the area near the edges of the ranges in which $c$ varies and therefore
are somewhat dependent on the particular regularization scheme. Still,
this dependence turns out to be weaker than what one would have been
inclined to believe on the basis of the few simulations that had no
continuous tuning ability for $c$ built in. Thus, some of the stronger
``universality'' evident in eq.~\ref{eq:effact} also makes its way into the
triviality bound.

There is an approximate physical picture associated with the models
we studied.  When the regularized model is nonlinear one has to think
about the Higgs resonance as a loose bound state of two pions in an
$I=0$, $J=0$ state. Pions in such a state attract because superposing
the field configurations corresponding to individual pions makes the
state look more like the vacuum and hence lowers the energy. The four
derivative term in the action can add or subtract to this attraction.
We found that the smallest cutoff effects are obtained when the
coupling $b_1$ of the four derivative term is set so that the term
induces the maximal possible repulsion between the pions, postponing
the appearance of the Higgs resonance to higher energies.

\if \epsfpreprint Y
\begin{figure}[htb]
\epsfxsize=\columnwidth
\epsffile{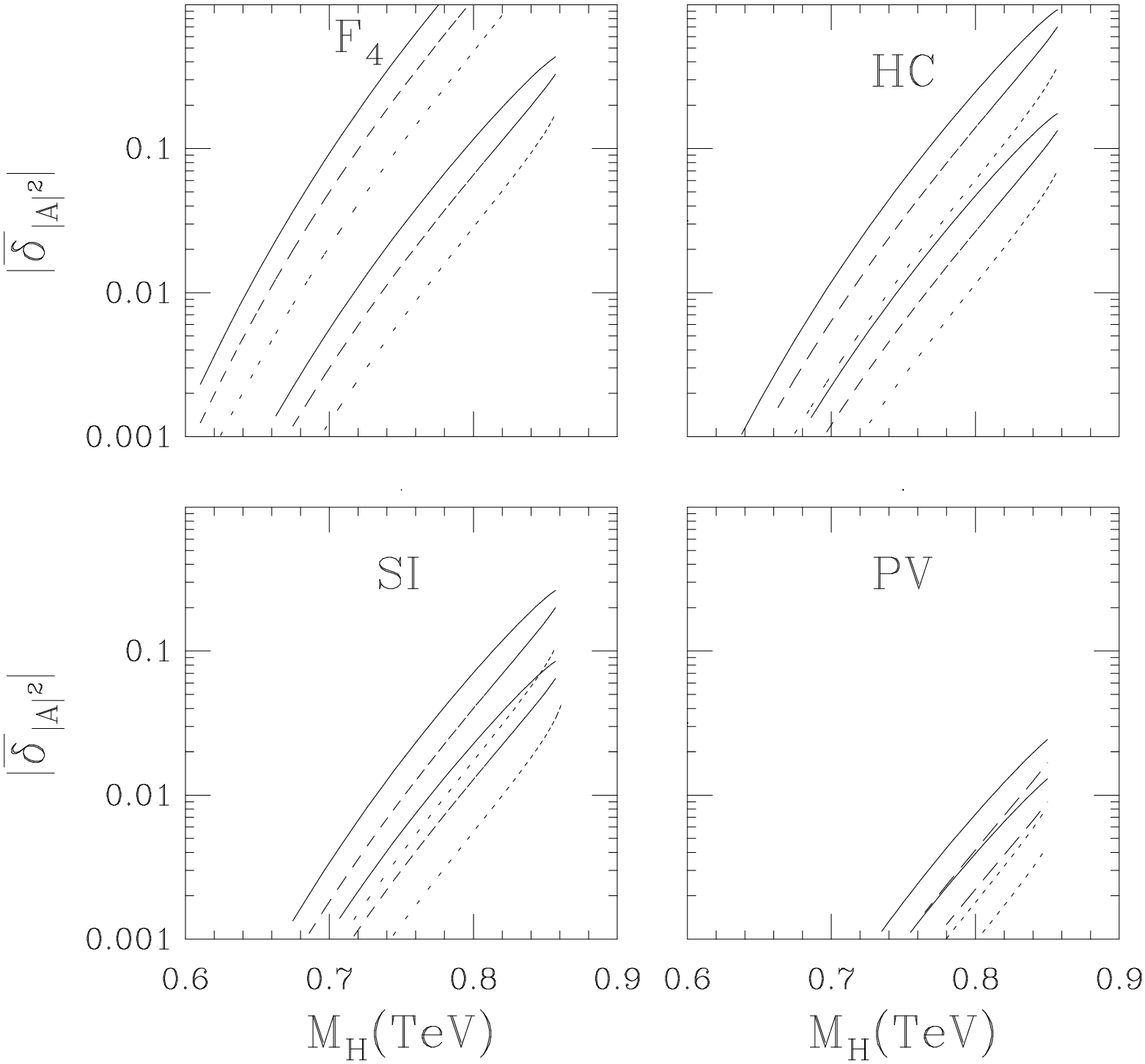}
\caption{\captionA}
\end{figure}
\fi

Our findings concerning the Higgs mass triviality bound are best summarized
in figure $1$ for the four different
regularization schemes considered. There the leading cutoff effects in
the cross section are plotted versus the Higgs mass in $TeV$
with the large $N$ results presented for $N=4$.
For all regularization
schemes we extract the bound by restricting the allowed cutoff
effects in the cross section to 4\%, at center of mass energies up to
twice the Higgs mass. From this figure it can be seen that turning
on the four derivative term to maximal allowed strength (maximum
repulsion between the pions) lowers the leading cutoff effects and
results to an approximately universal bound. Similar results are
obtained for the cutoff effects in the width. However these effects
are very small (less than $\approx 1\%$) and we therefore chose to
extract the Higgs mass bound using the more stringent cutoff effects in
the cross section.

\if \epsfpreprint Y
\begin{figure}[htb]
\epsfxsize=\columnwidth
\epsffile{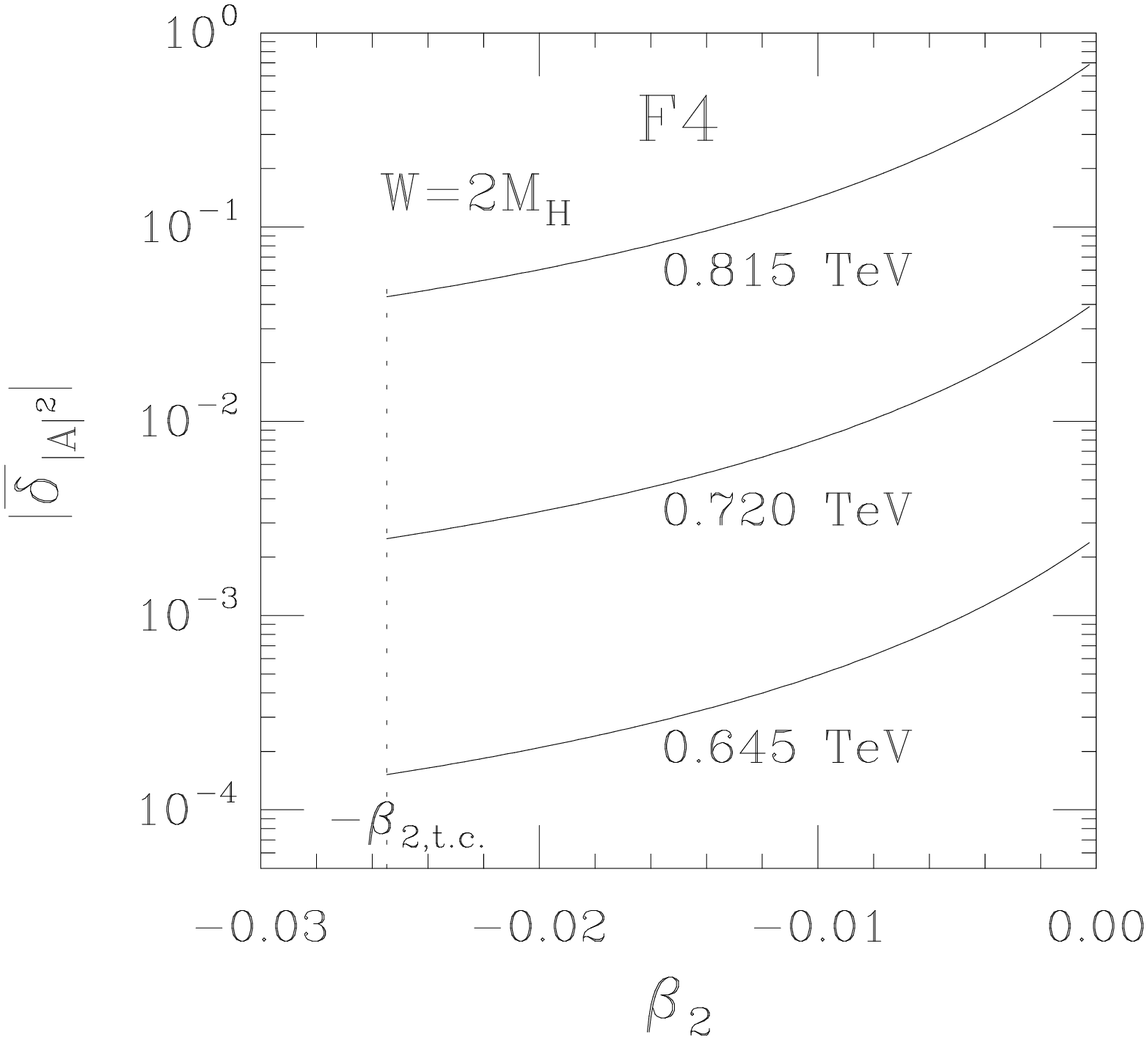}
\caption{\captionB}
\end{figure}
\fi

To see more clearly how the bound changes with the action we show an
example in figure $2$ for the $F_4$ lattice (the parameter
$\beta_2$ in the figure is proportional to $b_1$). Identical cutoff
effects, of 4{\%}, will be found at $N=\infty$ for a Higgs mass of
$0.720~TeV$ with the simplest action ($\beta_2=0$), and for a Higgs
mass of $0.815~TeV$ with an action that has the maximal amount of four
field interaction allowed ($\beta_2=-\beta_{2,t.c.}$). It is important
to understand that this difference is substantial when the width is
considered. In figure $3$ we show the regularization
independent part of the width as a function of the Higgs mass (note
that the leading order weak coupling approximation severely
underestimates the width when the Higgs mass becomes large). In our
example the width went from $0.320~TeV$ to about $0.500~TeV$ and the
heavier Higgs is definitely strongly interacting.  There is no doubt
therefore that, at least at infinite $N$, stopping the search for the
Higgs mass bound at the study of the simplest possible actions would
have been misleading.

\if \epsfpreprint Y
\begin{figure}[htb]
\epsfxsize=\columnwidth
\epsffile{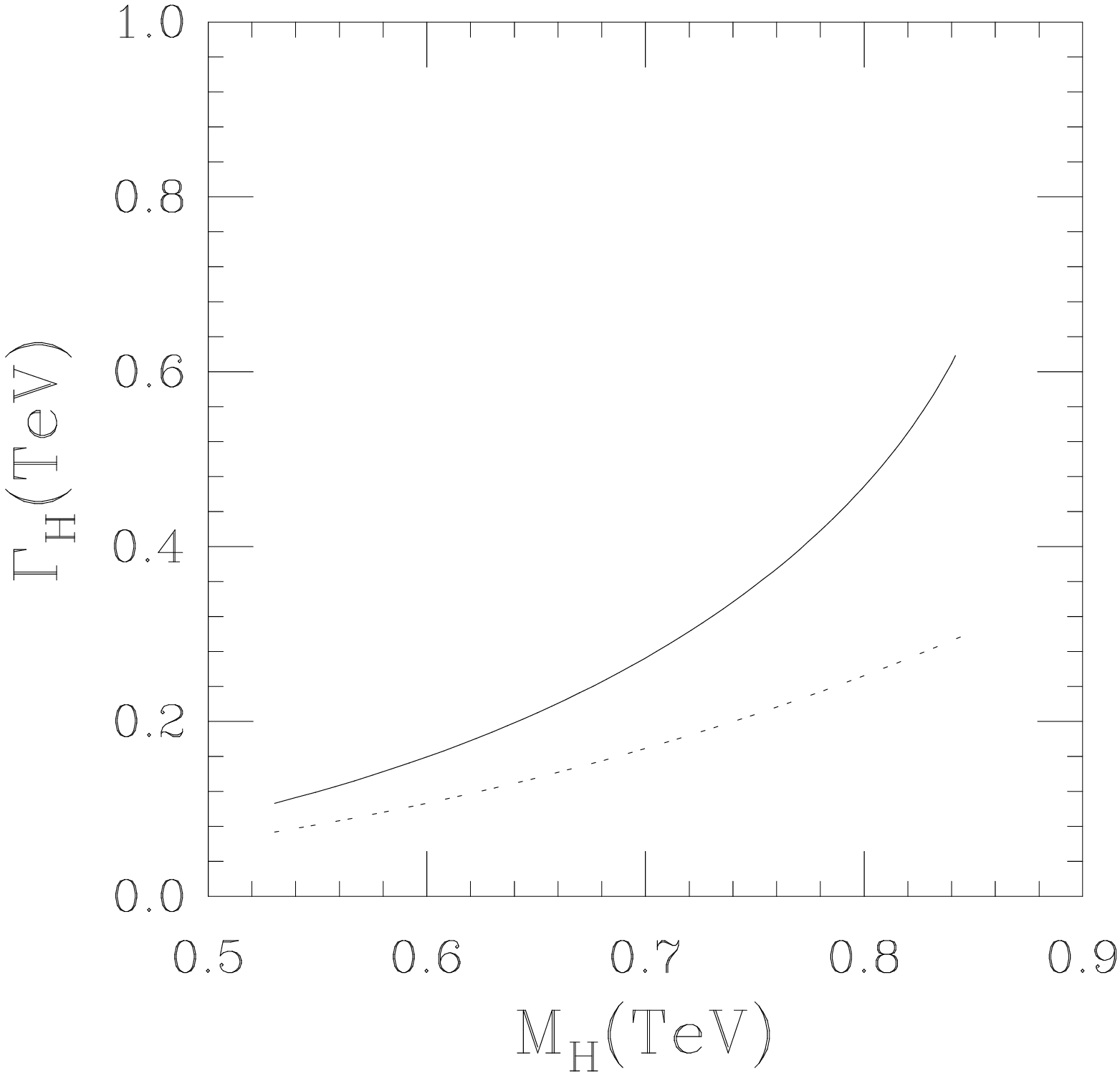}
\caption{\captionC}
\end{figure}
\fi

\addtocounter{footnote}{-1}
To make predictions for the physically relevant case $N=4$ we use the
known differences between the $N=\infty$ and the $N=4$ numerical
results, for the simplest lattice actions, to extrapolate to the
actions with four derivative terms that have not yet been studied
numerically. This extrapolation is sensible because the expansion in
$1/N$ is expected to be ``well behaved'' in the region where the
triviality bound is obtained.\footnote{The beta function to one loop
is proportional to $1+8/N$ and thus the $N=4$ answer is larger by a
factor of $3$ than the $N=\infty$ answer. However, this would make the
$N=\infty$ predictions unreliable only for quantities calculated at
energies much lower than the cutoff. The calculation of the Higgs mass
bound involves processes at energies close to the cutoff and therefore
there are not two very different scales to be connected and the beta
function is not needed.} We find that the large $N$ results
consistently overestimate the $N=4$ results by roughly $1/N=25\%$.

Based on our large $N$ results and using these extrapolations,
it seems that a more realistic and not
overly conservative estimate for the Higgs mass triviality bound is
$0.750~TeV$, and not $0.650~TeV$ as it is sometimes stated. At
$0.750~TeV$ the Higgs particle is expected to have a width of about
$0.290~TeV$ (after subtracting a possible overestimation of about
$25\%$) and is therefore quite strongly interacting. Such a heavy and
broad Higgs may be hard to detect experimentally.

\section{Numerical analysis of the $N=4$ case}

\if \epsfpreprint Y
\begin{figure}[t]
\epsfxsize=\columnwidth
\epsffile{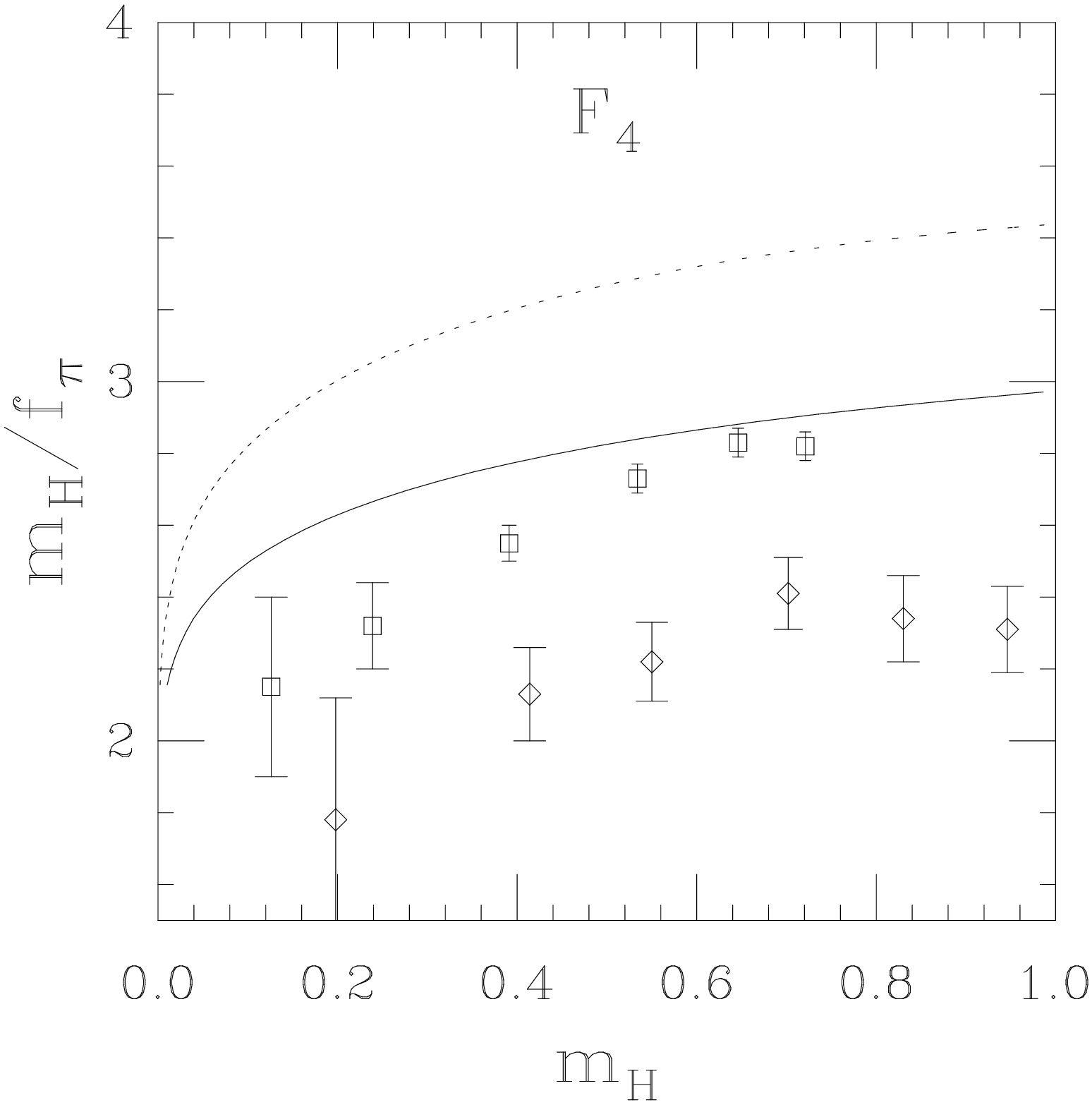}
\caption{\captionD}
\end{figure}
\fi

The physical $N=4$ case has to be studied numerically. We choose to
perform the simulation on the $F_4$ lattice because it does not
introduce lattice artifacts to the order we are interested in (order
$1/\Lambda^2$ where $\Lambda$ is the cutoff)
\cite{{BBHN1},{PLBF4},{BBHN2}}. We use a single cluster Wolff type
algorithm, and we measure the Higgs mass from the decay of the time
slice correlation function of the zero spatial momentum Higgs field.
In figure $4$ we present our large $N$ and numerical
results for the action with the four derivative term turned on to
maximum allowed strength. For comparison we have included in the same
figure the large $N$ and numerical~\cite{BBHN2} results for the naive
action. It can be seen from this figure that, for a given value of the
Higgs mass in lattice units ($m_H$), the Higgs mass in units of
$f_\pi=0.246~TeV$ increases approximately in agreement with the large
$N$ prediction when the four derivative term is turned on. This
supports the conclusions of the previous section.

For the $16^4$ lattice the largest Higgs mass in lattice units is
$\approx 0.70$ and it is therefore slightly larger than the energy of
the lowest free two pion $I=0$, $J=0$ state which is $\approx 0.68$.
It is therefore possible that the lowest lying two pion $I=0$, $J=0$
state may ``mix'' with the Higgs particle state. In that event the
extraction of the Higgs mass from the time slice correlation function
may become unreliable.

\if \epsfpreprint Y
\begin{figure}[t]
\epsfxsize=\columnwidth
\epsffile{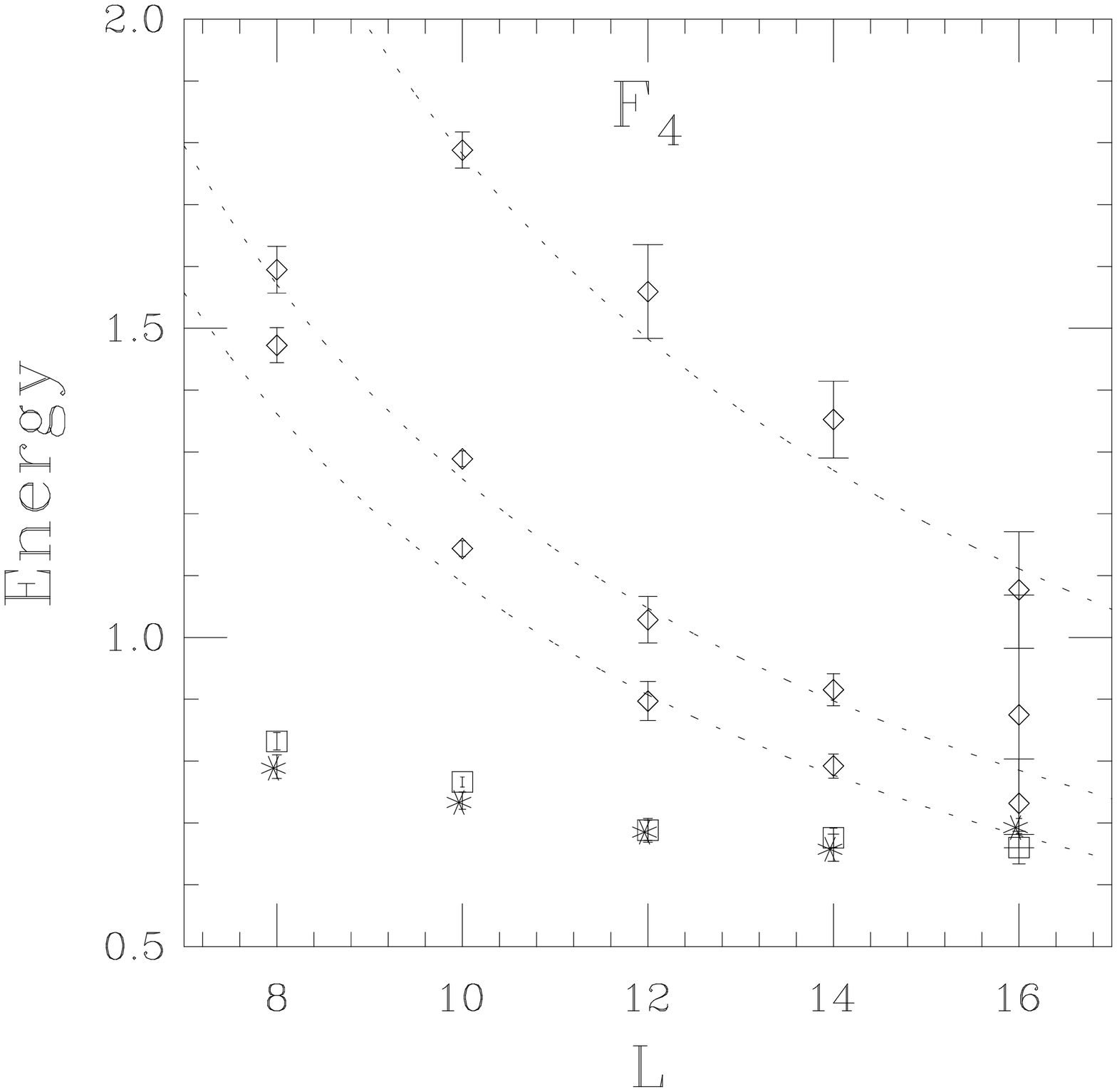}
\caption{\captionE}
\end{figure}
\fi

To explore this problem we look at the low lying spectrum in that
channel using the technique of reference~\cite{LW}. The results, with the
couplings set so that the largest Higgs mass with the
smallest possible cutoff effects is obtained, are
shown in figure $5$. The x-axis is the lattice size and
the y-axis is the energy in lattice units of the various states.
The dotted lines describe the low lying free
two pion $I=0$, $J=0$ states. The diamonds are the numerical values
for the two pion $I=0$, $J=0$ states and the squares are the numerical
values for the Higgs particle state. The stars are the numerical
values for the Higgs particle state obtained by simply using the decay
of the the time slice correlation function of the zero spatial
momentum Higgs field. As can be seen, the stars and the boxes are
the same, within errors, indicating that for our choice of operators
the ``mixing'' is very weak. Therefore, the Higgs mass extracted
from the decay of the the time slice correlation function
of the zero spatial momentum Higgs field is reliable and the results in
figure $4$ can be trusted.

\if \epsfpreprint N
\section{Figure captions}

\noindent{\bf Figure~$1$:~}\captionA

\noindent{\bf Figure~$2$:~}\captionB

\noindent{\bf Figure~$3$:~}\captionC

\noindent{\bf Figure~$4$:~}\captionD

\noindent{\bf Figure~$5$:~}\captionE
\fi

\vfill

\end{document}